\documentclass[pra,superscriptaddress,12pt,noshowpacs]{revtex4}
\usepackage[portuguese]{babel}
\usepackage[T1]{fontenc}
\usepackage[utf8]{inputenc}
\usepackage{graphicx,epstopdf}
\usepackage{amsmath}
\usepackage[pdftex,plainpages=false,colorlinks=true,citecolor=blue,linkcolor=blue,urlcolor=blue,filecolor=green,bookmarksopen=true]{hyperref}
\usepackage{amsfonts}
\usepackage{bbm}
\usepackage{amssymb}
\usepackage{color}
\usepackage{latexsym}
\usepackage{times,txfonts}
\usepackage{caption}
\usepackage{slashed}

\usepackage{dcolumn}
\usepackage{bm}

\newcommand{\be}{\begin{equation}}
\newcommand{\ee}{\end{equation}}
\newcommand{\bea}{\begin{eqnarray}}
\newcommand{\eea}{\end{eqnarray}}

\begin{document}

\title{One-loop Euler-Heisenberg action in Lorentz-violating QED revisited}

\author{R. Ara\'ujo}
\affiliation{Universidade Federal de Alagoas, 57072-900, Macei\'o, Alagoas, Brazil}
\email{raline.araujo,tmariz@fis.ufal.br}

\author{T. Mariz}
\affiliation{Universidade Federal de Alagoas, 57072-900, Macei\'o, Alagoas, Brazil}
\email{raline.araujo,tmariz@fis.ufal.br}

\author{J. R. Nascimento}
\affiliation{Departamento de F\'\i sica, Universidade Federal da Para\'\i ba,\\
 Caixa Postal 5008, 58051-970, Jo\~ao Pessoa, Para\'\i ba, Brazil}
\email{jroberto,petrov@fisica.ufpb.br}

\author{A. Yu. Petrov}
\affiliation{Departamento de F\'\i sica, Universidade Federal da Para\'\i ba,\\
 Caixa Postal 5008, 58051-970, Jo\~ao Pessoa, Para\'\i ba, Brazil}
\email{jroberto,petrov@fisica.ufpb.br}

\date{\today}

\begin{abstract}
We discuss applications of the proper-time method in a Lorentz-violating extension of QED characterized by the addition of the term proportional to the antisymmetric tensor $H_{\mu\nu}$. Unlike other LV extensions of QED, in our case, the one-loop Euler-Heisenberg-like action turns out to include only odd powers of the stress tensor $F_{\mu\nu}$. Our result is shown to be UV finite, and it is confirmed using the Feynman diagrams framework. 
\end{abstract}

\maketitle

\section{Introduction}

An important line of studies of Lorentz-violating (LV) theories consists of obtaining the quantum corrections in these theories. The first example of such a study was performed already in the seminal paper \cite{Colladay} where the LV Standard Model extension (LV SME) was formulated. Various studies of quantum corrections in LV theories have been performed, where the most important ones are renormalization of LV parameters in LV SME (see e.g. \cite{KosPic}) and calculation of finite quantum corrections in LV theories, which is naturally treated as a perturbative generation of LV terms (interesting examples of such calculations are given e.g. in \cite{JK,aether}, and a review on obtaining finite corrections to various LV terms is presented in \cite{ourLV}).

In this context, one of the interesting problems in studying the effective action in various extensions of QED is certainly to obtain the one-loop Euler-Heisenberg (EH) effective action \cite{EH}, which involves all orders in the stress tensor $F_{\mu\nu}$. It has been treated in many contexts (for a review, see \cite{Dunne}). Therefore, computing it in LV theories is a rather natural task. The first examples of such calculations are presented in \cite{EH1} for certain minimal LV extensions of the spinor QED, and in \cite{scalpt} for LV scalar QED. However, only a few examples of LV extensions of QED have been studied up to now within the EH context. Explicitly, for the spinor LV QED, the LV terms proportional to $b^{\mu}$ and $c^{\mu\nu}$ have been studied in \cite{EH1}. At the same time, the LV term proportional to $a_{\mu}$ is easily shown to yield a trivial result (these LV tensors are defined in \cite{KosPic} and listed below in (\ref{genrenmod})). So, it is natural to consider others. Within this paper, we concentrate on the impacts of the LV CPT-even additive term $\frac{1}{2}\bar{\psi}H^{\mu\nu}\sigma_{\mu\nu}\psi$ in the Lagrangian, for which we will find the EH effective action. This is the aim we pursue in this paper. 

The structure of the paper is as follows. In Section \ref{pt}, we calculate the EH effective action to the first order in $H_{\mu\nu}$. in Section \ref{wf}, we study the three-point function of the gauge field through the Feynman diagrams framework and explicitly demonstrate that it matches the result that can be read off from the EH effective action. Finally, in Section \ref{su}, the discussion of our results is presented.

\section{Proper-time method}\label{pt}

The spinor sector of the minimal LV QED is described by the Lagrangian \cite{KosPic}
\begin{equation}
	{\cal L}=\bar{\psi}(i\Gamma^{\nu}D_{\nu}-M)\psi,\label{genrenmod}
\end{equation}
where $D_\mu=\partial_\mu+ieA_\mu$, with
\begin{eqnarray}
	\Gamma^{\nu} &=&\gamma^{\nu}+c^{\mu\nu}\gamma_{\mu}+d^{\mu\nu}\gamma_{\mu}\gamma_{5}+e^{\nu}+if^{\nu}\gamma_{5}+{\textstyle\frac12}g^{\lambda\mu\nu}\sigma_{\lambda\mu},\nonumber \\
	M &=&m+a_{\mu}\gamma^{\mu}+b_{\mu}\gamma^{\mu}\gamma_{5}+{\textstyle\frac12}H^{\mu\nu}\sigma_{\mu\nu}.
\end{eqnarray}
Our aim consists in studying the one-loop low-energy effective action in the gauge sector, i.e., the LV EH action. Effectively, we must obtain the contribution to the effective action of the lower order in a corresponding LV parameter, but including all orders in fields, from the following functional determinant
\bea
\label{fundet}
\Gamma^{(1)}&=&-i{\rm Tr}\ln (i\Gamma^{\nu}D_{\nu}-M).
\eea
Here and further, ${\rm Tr}$ is for the functional trace, while ${\rm tr}$ is for the simple matrix trace.
In our earlier studies \cite{EH1}, we have obtained the complete one-loop low-energy effective action (LEEA), that is, the EH action, for the lower orders in $b_{\mu}$ and $c_{\mu\nu}$, while the impact of $a^{\mu}$ is trivial since the $a^{\mu}$ can be removed through a redefinition of a spinor field, see \cite{KosPic}. As argued, and as it is very natural, the lower order in $b_{\mu}$ for the EH action must be the second one (the first order will yield either CFJ term or higher-derivative (HD) terms like Myers-Pospelov or HD CFJ-like terms, see \cite{JK,HDLV}, which do not match the EH structure). It is clear also that the EH structure cannot be generated by the first orders in $e^{\mu}$, $f^{\mu}$, $g^{\mu\nu\lambda}$ since they have odd numbers of indices, and by the first order in $d^{\mu\nu}$ for parity reasons, since $d^{\mu\nu}$ is a pseudotensor, so that to yield an observer Lorentz scalar form the quantum correction must include even orders in this parameter.

Therefore, it is natural to study the possible one-loop effective action involving the first order in $H_{\mu\nu}$, which is expected to yield the EH form. While it is clear that the possible EH-like contributions of even orders in $F_{\mu\nu}$, like ${\rm tr}(HF^{2n})$ (e.g. $H_{\mu\nu}F^{\mu\alpha}F^{\nu}_{\phantom{\nu}\alpha}$), vanish by symmetry reasons, and $H^{\mu\nu}F_{\mu\nu}$ is a total derivative, nothing forbids the possibility of terms with higher odd orders in $F_{\mu\nu}$, like  ${\rm tr}(HF^{2n+1})$. So, let us study these contributions to the EH action.

We start with the expression (\ref{fundet}), which in our case takes the form
\bea
\label{fundet1}
\Gamma^{(1)}&=&-i{\rm Tr}\ln (i\Gamma^{\nu}D_{\nu}-M)=-i{\rm Tr}\ln (i\slashed{D}-m-{\textstyle\frac12}H^{\mu\nu}\sigma_{\mu\nu}).
\eea
To proceed with this expression, we perform the standard trick used e.g. in \cite{EH1}, that is, we split it into two parts and, in the latter, take $-{\textstyle\frac i2}{\rm Tr}\ln [(i\slashed{D}-m-{\textstyle\frac12}H^{\mu\nu}\sigma_{\mu\nu})\gamma_5^2]=-{\textstyle\frac i2}{\rm Tr}\ln [\gamma_5(-i\slashed{D}-m-{\textstyle\frac12}H^{\mu\nu}\sigma_{\mu\nu})\gamma_5]$, to deal with the operator of the second order in derivatives. Hence, we must find
  \bea
\label{fundet2}
\Gamma^{(1)}&=&-\frac{i}{2}{\rm Tr}\ln [(i\slashed{D}-m-{\textstyle\frac12}H^{\mu\nu}\sigma_{\mu\nu})(-i\slashed{D}-m-{\textstyle\frac12}H^{\mu\nu}\sigma_{\mu\nu})] \nonumber\\
&=&-\frac{i}{2}{\rm Tr}\ln[D^2+{\textstyle\frac e2}F_{\mu\nu}\sigma^{\mu\nu}+2H_{\mu\nu}\gamma^\mu D^\nu+m^2+mH_{\mu\nu}\sigma^{\mu\nu}+{\textstyle\frac 14}(H_{\rho\sigma}\sigma^{\rho\sigma})^2],
\eea
where we took into account that $[D_{\mu},D_{\nu}]=ieF_{\mu\nu}$. Now, we expand this expression up to the first order in $H^{\mu\nu}$:
 \bea
\label{fundet3}
\Gamma^{(1)}&=&
-\frac{i}{2}{\rm Tr}[(2H_{\kappa\lambda}\gamma^\kappa D^\lambda+mH_{\rho\sigma}\sigma^{\rho\sigma})(D^2+{\textstyle\frac e2}F_{\mu\nu}\sigma^{\mu\nu}+m^2)^{-1}].
\eea

To obtain the inverse operator $(D^2+{\textstyle\frac e2}F_{\mu\nu}\sigma^{\mu\nu}+m^2)^{-1}$, we use the standard proper-time prescription $A^{-1}=-i\int_0^{\infty} ds\ e^{isA}$. So, we face a problem of finding $e^{isA}$. Explicitly, our one-loop effective action (\ref{fundet3}) is
\bea
\label{fundet4}
\Gamma^{(1)}&=&
-\frac{1}{2}{\rm Tr}\int_0^{\infty}ds\ e^{ism^2}(2H_{\kappa\lambda}\gamma^\kappa D^\lambda+mH_{\kappa\lambda}\sigma^{\kappa\lambda})\ e^{is(D^2+{\textstyle\frac e2}F_{\mu\nu}\sigma^{\mu\nu})}.
\eea
As usual, we disregard derivatives of $F_{\mu\nu}$. Thus, to proceed with calculations, we have
\bea
\label{fundet5}
\Gamma^{(1)}&=&
-\frac{1}{2}{\rm tr}\int d^4x \int_0^{\infty}ds\ e^{ism^2}(2H_{\kappa\lambda}\gamma^\kappa D^\lambda+mH_{\kappa\lambda}\sigma^{\kappa\lambda})\ e^{{\textstyle\frac{ies}2}F_{\mu\nu}\sigma^{\mu\nu}}e^{isD^2}\delta^4(x-x')|_{x=x'},
\eea
where we can use the key identity derived earlier in \cite{Schwinger:1951nm,Dittrich:1985yb,McA}:
\bea
\label{key}
&&e^{isD^2}\delta^4(x-x')|_{x=x'}=\langle x|e^{isD^2}|x'\rangle|_{x=x'} =-\frac{i}{16\pi^2s^2}{\rm det}^{1/2}\left(\frac{esF}{\sinh(esF)}\right),
\eea
here and further $F$ within the determinant is the matrix whose elements are given by $F_{\mu\nu}$.
Indeed, it is clear that the term proportional to $D^\lambda$, in (\ref{fundet4}), yields a zero contribution to the trace being of odd order in Dirac matrices. Therefore, we are left with
\bea
\label{fundet6}
\Gamma^{(1)}&=&
\frac{im}{32\pi^2}{\rm tr}\int d^4x \int_0^{\infty}\frac{ds}{s^2}\ e^{ism^2}H_{\kappa\lambda}\sigma^{\kappa\lambda}\ e^{{\textstyle\frac{ies}2}F_{\mu\nu}\sigma^{\mu\nu}} {\rm det}^{1/2}\left(\frac{esF}{\sinh(esF)}\right).
\eea
It remains to expand the exponential in a power series in $F_{\mu\nu}$:
\bea
\label{fundet7}
\Gamma^{(1)}&=&
\frac{im}{32\pi^2}{\rm tr}\int d^4x \int_0^{\infty}\frac{ds}{s^2}\ e^{ism^2} H_{\kappa\lambda}\sigma^{\kappa\lambda}\sum\limits_{n=0}^{\infty}\frac{({\textstyle\frac{ies}2}F_{\mu\nu}\sigma^{\mu\nu})^{2n+1}}{(2n+1)!} {\rm det}^{1/2}\left(\frac{esF}{\sinh(esF)}\right).
\eea
We note that only odd powers of $F_{\mu\nu}$ in the effective action (\ref{fundet7}) yield nontrivial contributions, since all objects like ${\rm tr}(HF^{2n})$ identically vanish.

It is instructive to calculate the second contribution of (\ref{fundet7}), namely the term proportional to ${\rm tr}(HF^{3})$, which corresponds to the Lorentz-violating EH action associated with the coefficient $H_{\mu\nu}$. To this end, we must also consider the expansion of the determinant, which reads
\begin{equation}
{\rm det}^{1/2}\left(\frac{esF}{\sinh(esF)}\right)=1+\frac{e^2s^2}{12}F_{\mu\nu}F^{\mu\nu}+{\cal O}(F^4).
\end{equation}
Thus, after evaluating the trace of a product of $\sigma^{\mu\nu}$ matrices in the term of third order in $F_{\mu\nu}$ from (\ref{fundet7}), we obtain
\bea
\Gamma_{EH}&=&
\frac{im}{32\pi^2}\int d^4x \int_0^{\infty}\frac{ds}{s^2}\ e^{ism^2} \nonumber\\
&&\times\left[\frac{ie^3s^3}{24}8H_{\kappa\lambda}F^{\kappa\lambda}F_{\mu\nu}F^{\mu\nu}-\frac{ie^3s^3}{48}(48H_{\kappa\lambda}F^{\kappa\lambda}F_{\mu\nu}F^{\mu\nu}-64H_{\kappa\lambda}F^{\lambda\mu}F_{\mu\nu}F^{\nu\kappa}) \right].
\eea
Finally, after performing the $s$-integration, we find
\begin{equation}
\label{3fields}
\Gamma_{EH} =  -\frac{e^3}{48 \pi ^2 m^3} \int d^4x\ (H_{\kappa\lambda}F^{\kappa\lambda}F_{\mu\nu}F^{\mu\nu} -2H_{\kappa\lambda}F^{\lambda\mu}F_{\mu\nu}F^{\nu\kappa}).
\end{equation}
In the next section, we check this result with use of the Feynman diagrams.

Now, to study the convergence of the series in (\ref{fundet7}), we can use the identity
\bea
\{\sigma^{\alpha\beta},\sigma^{\gamma\delta}\}=g^{\alpha\gamma}g^{\beta\delta}-g^{\alpha\delta}g^{\beta\gamma}+i\epsilon^{\alpha\beta\gamma\delta}\gamma_5
,\eea
which implies
\bea
(F_{\mu\nu}\sigma^{\mu\nu})^2=F_{\mu\nu}F^{\mu\nu}+iF_{\mu\nu}\tilde{F}^{\mu\nu}\gamma_5.
\eea
Hence, by also taking into account that the term with $n=0$ is trivial and can be disregarded, we can rewrite (\ref{fundet7}) as
\bea
\label{fundet8}
\Gamma^{(1)}&=&
\frac{im}{32\pi^2}{\rm tr}\int d^4x\int_0^{\infty}\frac{ds}{s^2}e^{ism^2} H_{\kappa\lambda}\sigma^{\kappa\lambda} \sum\limits_{n=1}^{\infty}\frac{({\textstyle\frac{ies}2})^{2n+1}}{(2n+1)!} (F_{\mu\nu}F^{\mu\nu}+iF_{\mu\nu}\tilde{F}^{\mu\nu}\gamma_5)^n\ F_{\rho\tau}\sigma^{\rho\tau} \nonumber\\&&\times\ {\rm det}^{1/2}\left(\frac{esF}{\sinh(esF)}\right).
\eea
Then, using the general formula
\begin{equation}
(a+b\gamma_5)^k=\frac{(a+b)^k+(a-b)^k}{2}+\gamma_5\frac{(a+b)^k-(a-b)^k}{2},
\end{equation}
valid for any positive integer $k$, and taking $a=F_{\mu\nu}F^{\mu\nu}$, and $b=iF_{\mu\nu}\tilde{F}^{\mu\nu}$, we obtain
\bea
\label{fundet9}
\Gamma^{(1)}&=&
\frac{im}{64\pi^2}{\rm tr}\int d^4x\int_0^{\infty}\frac{ds}{s^2}e^{ism^2} H_{\kappa\lambda}\sigma^{\kappa\lambda}F_{\rho\tau}\sigma^{\rho\tau} \sum\limits_{n=1}^{\infty} \frac{({\textstyle\frac{ies}2})^{2n+1}}{(2n+1)!} \nonumber\\ 
&&\times[(F_{\mu\nu}F^{\mu\nu}+iF_{\mu\nu}\tilde{F}^{\mu\nu})^{n}+(F_{\mu\nu}F^{\mu\nu}-iF_{\mu\nu}\tilde{F}^{\mu\nu})^{n}\nonumber\\ 
&&+\gamma_5(F_{\mu\nu}F^{\mu\nu}+iF_{\mu\nu}\tilde{F}^{\mu\nu})^{n}-\gamma_5(F_{\mu\nu}F^{\mu\nu}-iF_{\mu\nu}\tilde{F}^{\mu\nu})^{n}] \nonumber\\
&&\times\ {\rm det}^{1/2}\left(\frac{esF}{\sinh(esF)}\right).
\eea
It now remains to calculate the traces:
\bea
&&{\rm tr}(\sigma^{\mu\nu}\sigma^{\rho\sigma})=4(g^{\mu\rho}g^{\nu\sigma}-g^{\mu\sigma}g^{\nu\rho}),\nonumber\\
&&{\rm tr}(\sigma^{\mu\nu}\sigma^{\rho\sigma}\gamma_5)=-4i\epsilon^{\mu\nu\rho\sigma}.
\eea
So, we obtain
\bea
\label{fundet10}
\Gamma^{(1)}&=&
\frac{im}{16\pi^2}\int d^4x\int_0^{\infty}\frac{ds}{s^2}e^{ism^2}\sum\limits_{n=1}^{\infty} \frac{({\textstyle\frac{ies}2})^{2n+1}}{(2n+1)!} \nonumber\\ 
&&\times\{H_{\kappa\lambda}F^{\kappa\lambda}[(F_{\mu\nu}F^{\mu\nu}+iF_{\mu\nu}\tilde{F}^{\mu\nu})^{n}+(F_{\mu\nu}F^{\mu\nu}-iF_{\mu\nu}\tilde{F}^{\mu\nu})^{n}] \nonumber\\ 
&&-i\epsilon^{\kappa\lambda\rho\tau}H_{\kappa\lambda}F_{\rho\tau}
[(F_{\mu\nu}F^{\mu\nu}+iF_{\mu\nu}\tilde{F}^{\mu\nu})^{n}-(F_{\mu\nu}F^{\mu\nu}-iF_{\mu\nu}\tilde{F}^{\mu\nu})^{n}]\} \nonumber\\
&&\times\ {\rm det}^{1/2}\left(\frac{esF}{\sinh(esF)}\right).
\eea
This sum can be evaluated. We note that the result is finite (indeed, the sum begins with $s^3$) as it must be for dimensional reasons. 

To perform the sum, we introduce the notation: $\Sigma=F_{\mu\nu}F^{\mu\nu}+iF_{\mu\nu}\tilde{F}^{\mu\nu}$, $\bar{\Sigma}=F_{\mu\nu}F^{\mu\nu}-iF_{\mu\nu}\tilde{F}^{\mu\nu}$. So, we can write
\bea
\label{fundet11}
\Gamma^{(1)}&=&
\frac{im}{16\pi^2}\int d^4x\int_0^{\infty}\frac{ds}{s^2}e^{ism^2}\sum\limits_{n=1}^{\infty} \frac{({\textstyle\frac{ies}2})^{2n+1}}{(2n+1)!}
[H_{\kappa\lambda}F^{\kappa\lambda}(\Sigma^{n}+\bar{\Sigma}^{n})-i\epsilon^{\kappa\lambda\rho\tau}H_{\kappa\lambda}F_{\rho\tau}(\Sigma^{n}-\bar{\Sigma}^{n})]\nonumber\\
&&\times\ {\rm det}^{1/2}\left(\frac{esF}{\sinh(esF)}\right).
\eea
Using the Taylor series $\sum\limits_{n=1}^{\infty}\frac{A^n}{(2n+1)!}=\frac{\sinh \sqrt{A}-\sqrt{A}}{\sqrt{A}}$, we finally obtain 
\bea
\label{fundet12}
\Gamma^{(1)}&=&
-\frac{em}{32\pi^2}\int d^4x\int_0^{\infty}\frac{ds}{s}e^{ism^2}
\left[H_{\kappa\lambda}F^{\kappa\lambda}\left(\frac{\sinh({\textstyle\frac{ies}2}\sqrt{\Sigma})-{\textstyle\frac{ies}2}\sqrt{\Sigma}}{{\textstyle\frac{ies}2}\sqrt{\Sigma}}+\frac{\sinh({\textstyle\frac{ies}2}\sqrt{\bar{\Sigma}})-{\textstyle\frac{ies}2}\sqrt{\bar{\Sigma}}}{{\textstyle\frac{ies}2}\sqrt{\bar{\Sigma}}}\right)\right. \\
&&\left.+i\epsilon^{\kappa\lambda\rho\tau}H_{\kappa\lambda}F_{\rho\tau}\left(\frac{\sinh({\textstyle\frac{ies}2}\sqrt{\Sigma})-{\textstyle\frac{ies}2}\sqrt{\Sigma}}{{\textstyle\frac{ies}2}\sqrt{\Sigma}}-\frac{\sinh({\textstyle\frac{ies}2}\sqrt{\bar{\Sigma}})-{\textstyle\frac{ies}2}\sqrt{\bar{\Sigma}}}{{\textstyle\frac{ies}2}\sqrt{\bar{\Sigma}}}\right)
\right]\ {\rm det}^{1/2}\left(\frac{esF}{\sinh(esF)}\right).\nonumber
\eea

This is our final result. As we already noted, it is real; moreover, actually, (\ref{fundet12}) is nothing more as a compact presentation of (\ref{fundet11}), which involves only integer degrees of $\Sigma$ and $\bar{\Sigma}$ (hence there is no ambiguities caused by a multi-valued nature of a square root of a complex number in this result), so, the (\ref{fundet12}) must be interpreted as a power series in $\Sigma$ and $\bar{\Sigma}$.  We note also that this result is UV finite and begins with the cubic order in the stress tensor $F_{\mu\nu}$, which is nontrivial since earlier, in \cite{EH1,scalpt}, only even orders in the stress tensor were shown to yield nontrivial contributions to the effective action.

\section{Three-point function}\label{wf}

In this section, we analyze the three-point function to calculate the one-loop EH effective action of the parameter $H_{\mu\nu}$ using the Feynman diagram approach. We aim to explicitly demonstrate that the result obtained reproduces the same structure that can be derived directly from the EH effective action (\ref{fundet7}).

For this, given the Lagrangian 
\begin{equation}
{\cal L}=\bar{\psi}(i\slashed{D}-m-{\textstyle\frac12}H^{\mu\nu}\sigma_{\mu\nu})\psi,
\end{equation}
we must expand the propagator
\begin{equation}
\frac{i}{\slashed{p}-m-{\textstyle\frac12}H^{\mu\nu}\sigma_{\mu\nu}}=\frac{i}{\slashed{p}-m}+\frac{i}{\slashed{p}-m}(-{\textstyle\frac i2}H^{\mu\nu}\sigma_{\mu\nu})\frac{i}{\slashed{p}-m}+\cdots,
\end{equation}
so that $-{\textstyle\frac i2}H^{\mu\nu}\sigma_{\mu\nu}$ is considered as an insertion into the propagator $iS(p)=i(\slashed{p}-m)^{-1}$. Then, the third-order $A_\mu$ effective action becomes
\begin{eqnarray}\label{Gamma13}
\Gamma^{(1,3)}&=&\frac{1}{3} \int d^4x \int d^4k_1 d^4k_2 d^4k_3\ e^{-i(k_1+k_2+k_3)\cdot x}  \frac{1}{4}G_H^{\mu_1\mu_2\mu_3}(k_1,k_2,k_3) A_{\mu_1}(k_1)A_{\mu_2}(k_2)A_{\mu_3}(k_3),
\end{eqnarray}
with
\begin{equation}\label{GH}
G_H^{\mu_1\mu_2\mu_3}(k_1,k_2,k_3) = 2T_H^{\mu_1\mu_2\mu_3}(k_1,k_2,k_3) +  2T_H^{\mu_2\mu_1\mu_3}(k_2,k_1,k_3),
\end{equation}
where, to first order in $H_{\mu\nu}$, both $T_H^{\mu_1\mu_2\mu_3}(k_1,k_2,k_3)$ and $T_H^{\mu_2\mu_1\mu_3}(k_2,k_1,k_3)$ can be decomposed into three diagrams. In particular, for $T_H^{\mu_1\mu_2\mu_3}(k_1,k_2,k_3)$, we have
\begin{equation}\label{TH}
T_H^{\mu_1\mu_2\mu_3}(k_1,k_2,k_3) = T_{H1}^{\mu_1\mu_2\mu_3}(k_1,k_2,k_3) +T_{H2}^{\mu_1\mu_2\mu_3}(k_1,k_2,k_3) +T_{H3}^{\mu_1\mu_2\mu_3}(k_1,k_2,k_3),
\end{equation}
with
\begin{subequations}\label{TH1TH2TH3}
\begin{eqnarray}
iT_{H1}^{\mu_1\mu_2\mu_3}(k_1,k_2,k_3) &=& -(-{\textstyle\frac i2})(-ie)^3 \int \frac{d^4p}{(2\pi)^4} \mathrm{tr}\ iS(p)H_{\alpha\beta}\sigma^{\alpha\beta}iS(p)\gamma^{\mu_1}iS(p_{1})\gamma^{\mu_2}iS(p_{12})\gamma^{\mu_3},\
\end{eqnarray}
\begin{eqnarray}
iT_{H2}^{\mu_1\mu_2\mu_3}(k_1,k_2,k_3) &=& -(-{\textstyle\frac i2})(-ie)^3 \int \frac{d^4p}{(2\pi)^4} \mathrm{tr}\ iS(p)\gamma^{\mu_1}iS(p_{1})H_{\alpha\beta}\sigma^{\alpha\beta}iS(p_{1})\gamma^{\mu_2}iS(p_{12})\gamma^{\mu_3},\
\end{eqnarray}
\begin{eqnarray}
iT_{H3}^{\mu_1\mu_2\mu_3}(k_1,k_2,k_3) &=& -(-{\textstyle\frac i2})(-ie)^3 \int \frac{d^4p}{(2\pi)^4} \mathrm{tr}\ iS(p)\gamma^{\mu_1}iS(p_{1})\gamma^{\mu_2}iS(p_{12})H_{\alpha\beta}\sigma^{\alpha\beta}iS(p_{12})\gamma^{\mu_3},\
\end{eqnarray}
\end{subequations}
where $p_1^\mu=p^\mu-k_1^\mu$ and  $p_{12}^\mu=p^\mu-k_1^\mu-k_2^\mu$. The corresponding Feynman diagrams for expressions~(\ref{GH}) and~(\ref{TH}) are shown in Figs.~\ref{diagrams1} and~\ref{diagrams2}, respectively, in the Appendix. It is easy to see that $T_{H2}^{\mu_1\mu_2\mu_3}$ and $T_{H3}^{\mu_1\mu_2\mu_3}$ can be obtained from $T_{H1}^{\mu_1\mu_2\mu_3}$ by performing the cyclic interchanges:
\begin{subequations}\label{TH2TH3}
\begin{eqnarray}
T_{H2}^{\mu_1\mu_2\mu_3}(k_1,k_2,k_3) &=& T_{H1}^{\mu_2\mu_3\mu_1}(k_2,k_3,k_1), \\
T_{H3}^{\mu_1\mu_2\mu_3}(k_1,k_2,k_3) &=& T_{H1}^{\mu_3\mu_1\mu_2}(k_3,k_1,k_2).
\end{eqnarray}
\end{subequations}

Therefore, we must focus our attention on the graph $T_{H1}^{\mu_1\mu_2\mu_3}$, in which, by first considering the Feynman parameterization, we obtain
\begin{eqnarray}
T_{H1}^{\mu_1\mu_2\mu_3}(k_1,k_2,k_3) &=& \int_0^1 dx_1 \int_0^{1-x_1} dx_2 \int \frac{d^4p}{(2\pi)^4} \frac{3ixe^3}{(p^2-M^2)^4} \nonumber\\
&&\times \mathrm{tr}\ (\slashed{q}+m)H_{\alpha\beta}\sigma^{\alpha\beta}(\slashed{q}+m)\gamma^{\mu_1}(\slashed{q}_1+m)\gamma^{\mu_2}(\slashed{q}_{12}+m)\gamma^{\mu_3},
\end{eqnarray}
where
\begin{equation}
M^2=m^2-x_1(1-x_1)k_1^2-x_{12}(1-x_{12})k_2^2-2x_1(1-x_{12})k_1\cdot k_2
\end{equation}
and
\begin{eqnarray}
q^\mu &=& p^\mu +(1-x_1)k_1^\mu +(1-x_{12})k_2^\mu, \\
q_1^\mu &=& p_1^\mu +(1-x_1)k_1^\mu +(1-x_{12})k_2^\mu,
\end{eqnarray}
with $x_{12}=x_1+x_2$, and so on. Then, after we calculate the trace over Dirac matrices and the corresponding integrals, up to order $1/m^3$, we arrive at
\begin{equation}
T_{H1}^{\mu_1\mu_2\mu_3}(k_1,k_2,k_3) =  \sum_{i=1}^{4} T_{H1ggi}^{\mu_1\mu_2\mu_3}(k_1,k_2,k_3) +\sum_{i=1}^{4} T_{H1gki}^{\mu_1\mu_2\mu_3}(k_1,k_2,k_3)  +\sum_{i=1}^{4} T_{H1kki}^{\mu_1\mu_2\mu_3}(k_1,k_2,k_3), 
\end{equation}
where the explicit form of all relevant terms is given in Eq.~(\ref{TH1}) (see the Appendix). In this expression, we use the following notation: $T^{\mu_1\mu_2\mu_3}_{H1ggi}$ denotes terms in which all external fields are contracted with Minkowski metrics; in $T^{\mu_1\mu_2\mu_3}_{H1gki}$, two of the external fields are contracted with Minkowski metrics and the remaining one with an external momentum; finally, in $T^{\mu_1\mu_2\mu_3}_{H1kki}$ only one external field is contracted with a Minkowski metric and the remaining two with external momenta. In all these expressions, $i = 1, 2, 3, 4$ labels the corresponding contributions. Analogous definitions are also employed in the next equation.

Let us now briefly discuss the question of photon double splitting in the collinear limit. If the incident on-shell photon has energy $E_1$ and momentum $\vec k_1$, then energy–momentum conservation requires that all momenta $\vec k_i$ must be aligned. Thus, the initial and final photons propagate collinearly, with orthogonal four-momenta $k_i^\mu k_{j\mu}=0$ (for more details, see \cite{Kostelecky:2002ue}). From the transversality condition, one finds  $\epsilon_{i\mu}k_j^\mu=0$ (or $A_\mu(k_i) k_j^\mu=0$). Hence, the contribution (\ref{H1gg1}) of order $\frac{1}{m}$ can yield a nonzero amplitude, while all other terms of (\ref{TH1}) of order $\frac{1}{m^3}$ vanish in this limit. However, when we consider the other contributions coming from $T_{H2}^{\mu_1\mu_2\mu_3}$ and $T_{H3}^{\mu_1\mu_2\mu_3}$ through the permutations (\ref{TH2TH3}), all the contributions of order $\frac{1}{m}$ cancel each other, and thus photon double splitting does not occur. This fact was only mentioned in \cite{Kostelecky:2002ue}, but not explicitly shown. 

For the photon triple splitting initially discussed in \cite{Kostelecky:2002ue}, and later in \cite{FuMa}, the contributions of order $\frac{1}{m^2}$ do not cancel each other when all the insertions of the coefficient $c_{\mu\nu}$ in the propagator are considered. Thus, after the collinear limit is taken into account, a nonzero amplitude is obtained, in contrast to what we observe here for the coefficient $H_{\mu\nu}$.

The next step is to analyze the generation of the Lorentz-violating EH action for the first-order correction. For this, we must calculate $G_H^{\mu_1\mu_2\mu_3}$ (Eq.~(\ref{GH})), by initially calculating $T_H^{\mu_1\mu_2\mu_3}$ (Eq.~(\ref{TH})), when we
interchange the uncontracted indices as well as the momentum indices for obtaining  $T_{H2}^{\mu_1\mu_2\mu_3}$ and $T_{H3}^{\mu_1\mu_2\mu_3}$ (Eq.~(\ref{TH2TH3})) from $T_{H1}^{\mu_1\mu_2\mu_3}$ (Eq.~(\ref{TH1})). Proceeding in this way, we obtain
\begin{equation}\label{GH2}
G_{H}^{\mu_1\mu_2\mu_3}(k_1,k_2,k_3) =  \sum_{i=1}^{3} G_{Hggi}^{\mu_1\mu_2\mu_3}(k_1,k_2,k_3) +\sum_{i=1}^{3} G_{Hgki}^{\mu_1\mu_2\mu_3}(k_1,k_2,k_3)  +\sum_{i=1}^{4} G_{Hkki}^{\mu_1\mu_2\mu_3}(k_1,k_2,k_3), 
\end{equation}
where the relevant terms are given in Eq.~(\ref{ghgg}) in the Appendix. We observe that the contributions of order $\frac{1}{m}$ vanish, as expected.

Thus, considering these results (\ref{GH2}), the effective action (\ref{Gamma13}), to first order in $H_{\mu\nu}$, takes the form
\begin{eqnarray}\label{GammaEH}
\Gamma_{EH}&=&\int d^4x \int d^4k_1 d^4k_2 d^4k_3\ e^{-i(k_1+k_2+k_3)\cdot x} G_H(k_1,k_2,k_3),
\end{eqnarray}
where
\begin{eqnarray}
G_H(k_1,k_2,k_3) &=& -\frac{ie^3}{144 \pi ^2 m^3} (H_{\kappa\lambda}F^{\kappa\lambda}(k_1)F_{\mu\nu}(k_2)F^{\mu\nu}(k_3) +H_{\kappa\lambda}F^{\kappa\lambda}(k_2)F_{\mu\nu}(k_1)F^{\mu\nu}(k_3) \nonumber\\
&& +H_{\kappa\lambda}F^{\kappa\lambda}(k_3)F_{\mu\nu}(k_1)F^{\mu\nu}(k_2) -2H_{\kappa\lambda}F^{\lambda\mu}(k_1)F_{\mu\nu}(k_2)F^{\nu\kappa}(k_3) \nonumber\\
&&-2F_{\kappa\lambda}(k_1)H^{\lambda\mu}F_{\mu\nu}(k_2)F^{\nu\kappa}(k_3) -2F_{\kappa\lambda}(k_1)F^{\lambda\mu}(k_2)H_{\mu\nu}F^{\nu\kappa}(k_3)),
\end{eqnarray}
with $F^{\mu\nu}(k_1)=k_1^\mu A^\nu(k_1)-k_1^\nu A^\mu(k_1)$, and so on. Then, inverting the Fourier transform in Eq.~(\ref{GammaEH}), the Lorentz-violating Euler-Heisenberg action becomes
\begin{equation}
\Gamma_{EH} =  -\frac{e^3}{48 \pi ^2 m^3} \int d^4x\ (H_{\kappa\lambda}F^{\kappa\lambda}F_{\mu\nu}F^{\mu\nu} -2H_{\kappa\lambda}F^{\lambda\mu}F_{\mu\nu}F^{\nu\kappa}),
\end{equation}
where now $F_{\mu\nu}=\partial_\mu A_\nu-\partial_\nu A_\mu$. We note that, using this definition of $F_{\mu\nu}$, this result can be easily rewritten in the form (\ref{Gamma13}), that is, in terms of the vector potential $A_{\mu}$. This result exactly coincides with the expression (\ref{3fields}) obtained above with the use of the proper time method, which confirms the validity of our calculations.

\section{Summary}\label{su}

We calculated the one-loop EH effective action for the LV spinor QED characterized by the LV parameter $H_{\mu\nu}$ of dimension one, which ensures finiteness of our result. Its unusual feature consists in the fact that, unlike the standard EH effective action (including many cases with the presence of LV terms, see \cite{EH1,scalpt}), our one-loop expression involves only odd orders in $F_{\mu\nu}$. Another important observation regarding our result is that it involves the first order in the LV parameter, while in many other studies of the EH effective action, the result is of the second order in corresponding LV parameters (see e.g. \cite{EH1,scalpt}). Therefore, our result may be the dominant contribution due to the well-known smallness of LV parameters \cite{datatables}. We explicitly demonstrated that the third-order result obtained through the EH framework coincides with that calculated through the Feynman diagrams approach. Effectively, we succeeded in obtaining results analogous to those found in  \cite{FuMa,Kostelecky:2002ue} for our theory. We plan to study phenomenological implications of our results in further studies.

Another interesting problem consists in obtaining the EH effective action depending on other minimal LV parameters presented in the model (\ref{genrenmod}), namely, $d^{\mu\nu},e^{\mu},f^{\mu},g^{\mu\nu\lambda}$. We note that these calculations will be more complicated than those performed here and in \cite{EH1}, since in the presence of these parameters, calculations of more involved traces will be necessary. We plan to perform these studies in our forthcoming papers.

{\bf Acknowledgments.}  The work of T. M. has been partially supported by the CNPq project No. 309360/2025-0 and FAPEAL project No. E:60030.0000002341/2022. The work of A. Yu.\ P. has been partially supported by the CNPq project No. 303777/2023-0. 

\section*{Appendix}

The relevant contributions to $T_{H1}^{\mu_1\mu_2\mu_3}(k_1,k_2,k_3)$ explicitly look like
\begin{subequations}\label{TH1}
\begin{eqnarray}\label{H1gg1}
T_{H1gg1}^{\mu_1\mu_2\mu_3}(k_1,k_2,k_3) &=& \frac{ie^3H_{\alpha\beta}}{48 \pi ^2 m} (g^{\mu _1 \mu _2} ((k_2^{\beta }-2 k_1^{\beta }) g^{\alpha  \mu _3}+(2 k_1^{\alpha }-k_2^{\alpha }) g^{\beta  \mu _3})+g^{\mu _1 \mu _3} (k_2^{\alpha } g^{\beta  \mu _2}-k_2^{\beta } g^{\alpha  \mu _2}) \nonumber\\
&&+g^{\mu _2 \mu _3} ((2 k_1^{\beta }+3 k_2^{\beta }) g^{\alpha  \mu _1}-(2 k_1^{\alpha }+3 k_2^{\alpha }) g^{\beta  \mu _1})),
\end{eqnarray}
\begin{eqnarray}
T_{H1gg2}^{\mu_1\mu_2\mu_3}(k_1,k_2,k_3) &=& -\frac{ie^3H_{\alpha\beta}}{480 \pi ^2 m^3} g^{\mu _1 \mu _2} (-2 k_1^{\alpha } (2 k_1^2+4 k_2^2+7 k_1\cdot k_2) g^{\beta  \mu _3}+k_2^{\alpha } (11 k_1^2+2 k_2^2+6 k_1\cdot k_2) g^{\beta  \mu _3} \nonumber\\
&&+g^{\alpha  \mu _3} (2 k_1^{\beta } (2 k_1^2+4 k_2^2+7 k_1\cdot k_2)-k_2^{\beta } (11 k_1^2+2 k_2^2+6 k_1\cdot k_2))),
\end{eqnarray}
\begin{eqnarray}
T_{H1gg3}^{\mu_1\mu_2\mu_3}(k_1,k_2,k_3) &=& -\frac{ie^3H_{\alpha\beta}}{480 \pi ^2 m^3} g^{\mu _1 \mu _3} (5 k_1^{\alpha } (k_2^2+2 k_1\cdot k_2) g^{\beta  \mu _2}-k_2^{\alpha } (11 k_1^2+2 k_2^2+6 k_1\cdot k_2) g^{\beta  \mu _2} \nonumber\\
&&+g^{\alpha  \mu _2} (k_2^{\beta } (11 k_1^2+2 k_2^2+6 k_1\cdot k_2)-5 k_1^{\beta } (k_2^2+2 k_1\cdot k_2))),
\end{eqnarray}
\begin{eqnarray}
T_{H1gg4}^{\mu_1\mu_2\mu_3}(k_1,k_2,k_3) &=& \frac{ie^3H_{\alpha\beta}}{480 \pi ^2 m^3} g^{\mu _2 \mu _3} (2 k_1^{\alpha } (-2 k_1^2+k_2^2+3 k_1\cdot k_2) g^{\beta  \mu _1}-5 k_2^{\alpha } (3 k_1^2+k_2^2+2 k_1\cdot k_2) g^{\beta  \mu _1} \nonumber\\
&&+g^{\alpha  \mu _1} (k_1^{\beta } (4 k_1^2-2 k_2^2-6 k_1\cdot k_2)+5 k_2^{\beta } (3 k_1^2+k_2^2+2 k_1\cdot k_2))),
\end{eqnarray}
\begin{eqnarray}
T_{H1gk1}^{\mu_1\mu_2\mu_3}(k_1,k_2,k_3) &=& -\frac{ie^3H_{\alpha\beta}}{48 \pi ^2 m} (g^{\alpha  \mu _1} ((k_2^{\mu _3}-2 k_1^{\mu _3}) g^{\beta  \mu _2}+3 (2 k_1^{\mu _2}+k_2^{\mu _2}) g^{\beta  \mu _3})-g^{\alpha  \mu _2} ((k_2^{\mu _3}-2 k_1^{\mu _3}) g^{\beta  \mu _1} \nonumber\\
&&+(2 k_1^{\mu _1}+3 k_2^{\mu _1}) g^{\beta  \mu _3})+g^{\alpha  \mu _3} ((2 k_1^{\mu _1}+3 k_2^{\mu _1}) g^{\beta  \mu _2}-3 (2 k_1^{\mu _2}+k_2^{\mu _2}) g^{\beta  \mu _1})),
\end{eqnarray}
\begin{eqnarray}
T_{H1gk2}^{\mu_1\mu_2\mu_3}(k_1,k_2,k_3) &=& \frac{ie^3H_{\alpha\beta}}{96 \pi ^2 m^3} (g^{\mu _1 \mu _2} (2 k_1^{\mu _3}+k_2^{\mu _3})-g^{\mu _1 \mu _3} (2 k_1^{\mu _2}+k_2^{\mu _2})+g^{\mu _2 \mu _3} (2 k_1^{\mu _1}+k_2^{\mu _1})) \nonumber\\
&&\times (k_2^{\alpha } k_1^{\beta }-k_1^{\alpha } k_2^{\beta }),
\end{eqnarray}
\begin{eqnarray}
T_{H1gk3}^{\mu_1\mu_2\mu_3}(k_1,k_2,k_3) &=& \frac{ie^3H_{\alpha\beta}}{480 \pi ^2 m^3} (g^{\alpha  \mu _1} (g^{\beta  \mu _2} (k_1^{\mu _3} (4 k_1^2+9 k_2^2+8 k_1\cdot k_2)-k_2^{\mu _3} (11 k_1^2+k_2^2+7 k_1\cdot k_2)) \nonumber\\
&&-5 g^{\beta  \mu _3} (k_1^{\mu _2} (3 k_2^2+4 (k_1^2+k_1\cdot k_2))+k_2^{\mu _2} (2 k_1^2+k_2^2+k_1\cdot k_2))) \nonumber\\
&&+g^{\alpha  \mu _2} g^{\beta  \mu _3} ((4 k_1^2+5 k_2^2) k_1^{\mu _1}+5 k_2^{\mu _1} (2 k_2^2+3 (k_1^2+k_1\cdot k_2)))),
\end{eqnarray}
\begin{eqnarray}
T_{H1gk4}^{\mu_1\mu_2\mu_3}(k_1,k_2,k_3) &=& -\frac{ie^3H_{\alpha\beta}}{480 \pi ^2 m^3} (g^{\alpha  \mu _2} g^{\beta  \mu _1} (k_1^{\mu _3} (4 k_1^2+9 k_2^2+8 k_1\cdot k_2)-k_2^{\mu _3} (11 k_1^2+k_2^2+7 k_1\cdot k_2)) \nonumber\\
&&+g^{\alpha  \mu _3} (g^{\beta  \mu _2} ((4 k_1^2+5 k_2^2) k_1^{\mu _1}+5 k_2^{\mu _1} (2 k_2^2+3 (k_1^2+k_1\cdot k_2))) \nonumber\\
&&-5 g^{\beta  \mu _1} (k_1^{\mu _2} (3 k_2^2+4 (k_1^2+k_1\cdot k_2))+k_2^{\mu _2} (2 k_1^2+k_2^2+k_1\cdot k_2)))),
\end{eqnarray}
\begin{eqnarray}
T_{H1kk1}^{\mu_1\mu_2\mu_3}(k_1,k_2,k_3) &=& -\frac{ie^3H_{\alpha\beta}}{60 \pi ^2 m^3} (k_1^{\mu _2} k_1^{\mu _1} ((k_1^{\beta }+k_2^{\beta }) g^{\alpha  \mu _3}-(k_1^{\alpha }+k_2^{\alpha }) g^{\beta  \mu _3})+k_1^{\mu _3} (k_1^{\mu _1} (k_2^{\alpha } g^{\beta  \mu _2}-k_2^{\beta } g^{\alpha  \mu _2}) \nonumber\\
&&+k_1^{\mu _2} (k_1^{\alpha } g^{\beta  \mu _1}-k_1^{\beta } g^{\alpha  \mu _1}))),
\end{eqnarray}
\begin{eqnarray}
T_{H1kk2}^{\mu_1\mu_2\mu_3}(k_1,k_2,k_3) &=& \frac{ie^3H_{\alpha\beta}}{480 \pi ^2 m^3} (k_2^{\mu _2} k_1^{\mu _1} ((k_2^{\beta }-4 k_1^{\beta }) g^{\alpha  \mu _3}+(4 k_1^{\alpha }-k_2^{\alpha }) g^{\beta  \mu _3})+k_2^{\mu _3} (k_1^{\mu _1} ((10 k_1^{\alpha }+k_2^{\alpha }) g^{\beta  \mu _2} \nonumber\\
&&-(10 k_1^{\beta }+k_2^{\beta }) g^{\alpha  \mu _2})+k_1^{\mu _2} ((14 k_1^{\beta }+5 k_2^{\beta }) g^{\alpha  \mu _1}-(14 k_1^{\alpha }+5 k_2^{\alpha }) g^{\beta  \mu _1}))),
\end{eqnarray}
\begin{eqnarray}
T_{H1kk3}^{\mu_1\mu_2\mu_3}(k_1,k_2,k_3) &=& -\frac{ie^3H_{\alpha\beta}}{480 \pi ^2 m^3} (k_1^{\mu _2} k_2^{\mu _1} ((6 k_1^{\alpha }+k_2^{\alpha }) g^{\beta  \mu _3}-(6 k_1^{\beta }+k_2^{\beta }) g^{\alpha  \mu _3})+k_1^{\mu _3} (k_2^{\mu _1} ((10 k_1^{\beta }+k_2^{\beta }) g^{\alpha  \mu _2} \nonumber\\
&&-(10 k_1^{\alpha }+k_2^{\alpha }) g^{\beta  \mu _2})+k_2^{\mu _2} ((5 k_2^{\beta }-4 k_1^{\beta }) g^{\alpha  \mu _1}+(4 k_1^{\alpha }-5 k_2^{\alpha }) g^{\beta  \mu _1}))),
\end{eqnarray}
\begin{eqnarray}
T_{H1kk4}^{\mu_1\mu_2\mu_3}(k_1,k_2,k_3) &=& -\frac{ie^3H_{\alpha\beta}}{240 \pi ^2 m^3} (2 k_2^{\mu _1} k_2^{\mu _2} ((k_1^{\alpha }+k_2^{\alpha }) g^{\beta  \mu _3}-(k_1^{\beta }+k_2^{\beta }) g^{\alpha  \mu _3})- k_2^{\mu _3} (k_2^{\mu _1} ((5 k_1^{\alpha }+2 k_2^{\alpha }) g^{\beta  \mu _2} \nonumber\\
&&-(5 k_1^{\beta }+2 k_2^{\beta }) g^{\alpha  \mu _2})+3 k_2^{\mu _2} (k_1^{\beta } g^{\alpha  \mu _1}-k_1^{\alpha } g^{\beta  \mu _1}))).
\end{eqnarray}
\end{subequations}
Then, the contributions to the effective action are given by
\begin{subequations}
\label{ghgg}
\begin{eqnarray}
G_{Hgg1}^{\mu_1\mu_2\mu_3}(k_1,k_2,k_3) &=& -\frac{ie^3H_{\alpha\beta}}{12 \pi ^2 m^3} g^{\mu _1 \mu _2} (k_1^{\alpha } (k_2^2-k_1\cdot k_2) g^{\beta  \mu _3}+k_2^{\alpha } (k_1^2-k_1\cdot k_2) g^{\beta  \mu _3}+g^{\alpha  \mu _3} \nonumber\\
&&\times (k_1^{\beta } (k_1\cdot k_2 -k_2^2) +k_2^{\beta } (k_1\cdot k_2-k_1^2))),
\end{eqnarray}
\begin{eqnarray}
G_{Hgg2}^{\mu_1\mu_2\mu_3}(k_1,k_2,k_3) &=& \frac{ie^3H_{\alpha\beta}}{12 \pi ^2 m^3} g^{\mu _1 \mu _3} (-(k_1^{\alpha } (k_2^2+2 k_1\cdot k_2) g^{\beta  \mu _2})+k_2^{\alpha } (2 k_1^2+k_1\cdot k_2) g^{\beta  \mu _2} \nonumber\\
&&+g^{\alpha  \mu _2} (k_1^{\beta } (k_2^2+2 k_1\cdot k_2)-k_2^{\beta } (2 k_1^2+k_1\cdot k_2))),
\end{eqnarray}
\begin{eqnarray}
G_{Hgg3}^{\mu_1\mu_2\mu_3}(k_1,k_2,k_3) &=& -\frac{ie^3H_{\alpha\beta}}{12 \pi ^2 m^3} g^{\mu _2 \mu _3} (-(k_1^{\alpha } (2 k_2^2+k_1\cdot k_2) g^{\beta  \mu _1})+k_2^{\alpha } (k_1^2+2 k_1\cdot k_2) g^{\beta  \mu _1} \nonumber\\
&&+g^{\alpha  \mu _1} (k_1^{\beta } (2 k_2^2+k_1\cdot k_2)-k_2^{\beta } (k_1^2+2 k_1\cdot k_2))),
\end{eqnarray}
\begin{eqnarray}
G_{Hgk1}^{\mu_1\mu_2\mu_3}(k_1,k_2,k_3) &=& \frac{ie^3H_{\alpha\beta}}{12 \pi ^2 m^3} (g^{\mu _1 \mu _2} (k_1^{\mu _3}-k_2^{\mu _3})-g^{\mu _1 \mu _3} (2 k_1^{\mu _2}+k_2^{\mu _2})+g^{\mu _2 \mu _3} (k_1^{\mu _1}+2 k_2^{\mu _1})) \nonumber\\
&&\times (k_2^{\alpha } k_1^{\beta }-k_1^{\alpha } k_2^{\beta }),
\end{eqnarray}
\begin{eqnarray}
G_{Hgk2}^{\mu_1\mu_2\mu_3}(k_1,k_2,k_3) &=& \frac{ie^3H_{\alpha\beta}}{12 \pi ^2 m^3} (g^{\alpha  \mu _1} (g^{\beta  \mu _2} (k_1^{\mu _3} (k_2^2+k_1\cdot k_2)-k_2^{\mu _3} (k_1^2+k_1\cdot k_2))+g^{\beta  \mu _3} \nonumber\\
&&\times (k_2^2 k_1^{\mu _2}-k_2^{\mu _2} k_1\cdot k_2))+g^{\alpha  \mu _2} g^{\beta  \mu _3} (k_1^2 k_2^{\mu _1}-k_1^{\mu _1} k_1\cdot k_2)),
\end{eqnarray}
\begin{eqnarray}
G_{Hgk3}^{\mu_1\mu_2\mu_3}(k_1,k_2,k_3) &=& -\frac{ie^3H_{\alpha\beta}}{12 \pi ^2 m^3} (g^{\alpha  \mu _2} g^{\beta  \mu _1} (k_1^{\mu _3} (k_2^2+k_1\cdot k_2)-k_2^{\mu _3} (k_1^2+k_1\cdot k_2))+g^{\alpha  \mu _3} \nonumber\\
&&\times (g^{\beta  \mu _1} (k_2^2 k_1^{\mu _2}-k_2^{\mu _2} k_1\cdot k_2)+g^{\beta  \mu _2} (k_1^2 k_2^{\mu _1}-k_1^{\mu _1} k_1\cdot k_2))),
\end{eqnarray}
\begin{eqnarray}
G_{Hkk1}^{\mu_1\mu_2\mu_3}(k_1,k_2,k_3) &=& -\frac{ie^3H_{\alpha\beta}}{12 \pi ^2 m^3} (k_1^{\mu _2} k_1^{\mu _1} (k_2^{\beta } g^{\alpha  \mu _3}-k_2^{\alpha } g^{\beta  \mu _3})+k_1^{\mu _3} (2 k_1^{\mu _1} (k_2^{\alpha } g^{\beta  \mu _2}-k_2^{\beta } g^{\alpha  \mu _2}) \nonumber\\
&&+k_1^{\mu _2} (k_2^{\beta } g^{\alpha  \mu _1}-k_2^{\alpha } g^{\beta  \mu _1}))),
\end{eqnarray}
\begin{eqnarray}
G_{Hkk2}^{\mu_1\mu_2\mu_3}(k_1,k_2,k_3) &=& -\frac{ie^3H_{\alpha\beta}}{12 \pi ^2 m^3} k_2^{\mu _3} (k_1^{\mu _1} (k_1^{\beta } g^{\alpha  \mu _2}-k_1^{\alpha } g^{\beta  \mu _2})+k_1^{\mu _2} ((k_2^{\beta }-k_1^{\beta }) g^{\alpha  \mu _1} \nonumber\\
&&+(k_1^{\alpha }-k_2^{\alpha }) g^{\beta  \mu _1})),
\end{eqnarray}
\begin{eqnarray}
G_{Hkk3}^{\mu_1\mu_2\mu_3}(k_1,k_2,k_3) &=& -\frac{ie^3H_{\alpha\beta}}{12 \pi ^2 m^3} (k_1^{\mu _2} k_2^{\mu _1} ((k_1^{\alpha }+k_2^{\alpha }) g^{\beta  \mu _3}-(k_1^{\beta }+k_2^{\beta }) g^{\alpha  \mu _3})+k_1^{\mu _3} (k_2^{\mu _1} ((k_1^{\beta }-k_2^{\beta }) g^{\alpha  \mu _2} \nonumber\\
&&+(k_2^{\alpha }-k_1^{\alpha }) g^{\beta  \mu _2})+k_2^{\mu _2} (k_2^{\beta } g^{\alpha  \mu _1}-k_2^{\alpha } g^{\beta  \mu _1}))),
\end{eqnarray}
\begin{eqnarray}
G_{Hkk4}^{\mu_1\mu_2\mu_3}(k_1,k_2,k_3) &=& \frac{ie^3H_{\alpha\beta}}{12 \pi ^2 m^3} (k_2^{\mu _2} k_2^{\mu _1} (k_1^{\alpha } g^{\beta  \mu _3}-k_1^{\beta } g^{\alpha  \mu _3})+k_2^{\mu _3} (k_2^{\mu _1} (k_1^{\alpha } g^{\beta  \mu _2}-k_1^{\beta } g^{\alpha  \mu _2}) \nonumber\\
&&+2 k_2^{\mu _2} (k_1^{\beta } g^{\alpha  \mu _1}-k_1^{\alpha } g^{\beta  \mu _1}))).
\end{eqnarray}
\end{subequations}

The corresponding Feynman diagrams are given below.

\vspace*{2mm}

\begin{figure}[htbp] 
	\begin{center} 
	\includegraphics[width={4cm}]{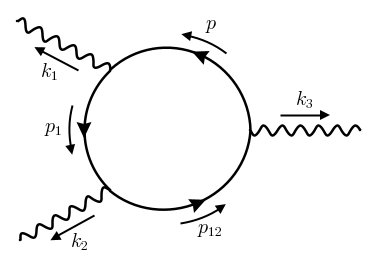}\hspace*{1cm} 
    \includegraphics[width={4cm}]{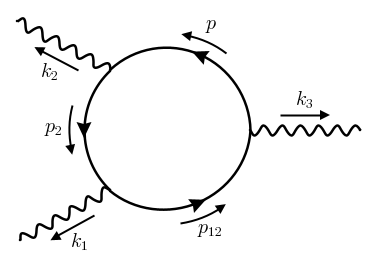}
	\end{center} 
	\caption{The two contributions to the three-point function.}
	\label{diagrams1} 
\end{figure}

\vspace*{2mm}

\begin{figure}[htbp] 
	\begin{center} 
	\includegraphics[width={4cm}]{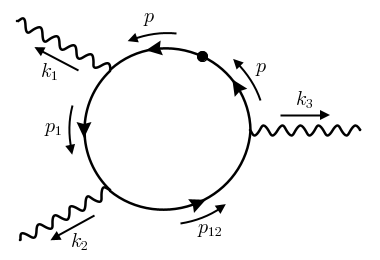}\hspace*{1cm} 
    \includegraphics[width={4cm}]{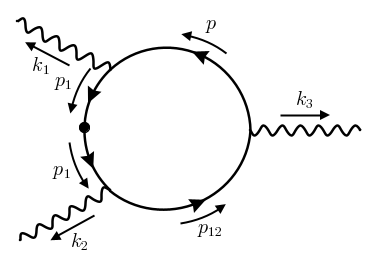}\hspace*{1cm} 
    \includegraphics[width={4cm}]{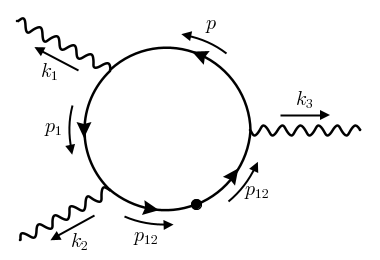}
	\end{center} 
	\caption{The first contribution with insertions of $H_{\mu\nu}$ in the propagator.}
	\label{diagrams2} 
\end{figure}

\vspace*{2mm}


\end{document}